\documentclass[a4paper]{article}
\usepackage{graphicx}
\usepackage{onecolpceurws}
\usepackage{wrapfig}
\usepackage{amsmath}
\usepackage{blindtext}
\usepackage{subfig}
\usepackage{esvect}
\usepackage[utf8]{inputenc}
\usepackage{color}
\usepackage[square,sort,comma,numbers]{natbib}

\usepackage[numbers]{natbib}

\title{Seismic-Net: A Deep Densely Connected Neural Network to Detect Seismic Events}

\author{Yue Wu, Youzuo Lin, Zheng Zhou, and Andrew Delorey}
\institution{Earth and Environment Sciences Division\\ Los Alamos National Laboratory\\
                Los Alamos, NM 87545 \\}

\begin{document}
\maketitle

\begin{abstract}
One of the risks of large-scale geologic carbon sequestration is the potential migration of fluids out of the storage formations. Accurate and fast detection of this fluids migration is not only important but also challenging, due to the large subsurface uncertainty and complex governing physics. Traditional leakage detection and monitoring techniques rely on geophysical observations
including seismic. However, the resulting accuracy of these methods is limited because of indirect information they provide requiring expert interpretation, therefore yielding in-accurate estimates of leakage rates and locations. In this work, we develop a novel machine-learning detection package, named``Seismic-Net'', which is based on the deep densely connected neural network. To validate the performance of our proposed leakage detection method, we employ our method to  a natural analog site at Chimayó, New Mexico. The seismic events in the data sets are generated because of the eruptions of geysers, which is due to the leakage of $\mathrm{CO}_\mathrm{2}$.  In particular, we demonstrate the efficacy of our Seismic-Net by formulating our detection problem as an event detection problem with time series data. A fixed-length window is slid throughout the time series data and we build a deep densely connected network to classify each window to determine if a geyser event is included. Through our numerical tests, we show that our model achieves precision/recall as high as 0.889/0.923. Therefore, our Seismic-Net has a great potential for detection of $\mathrm{CO}_\mathrm{2}$ leakage.
\end{abstract}
\vskip 32pt

\section{Introduction}

A critical issue for geologic carbon sequestration is the ability to detect the leakage of $\mathrm{CO}_\mathrm{2}$. There has been
three major different geophysical methods employed to detect the leakage of $\mathrm{CO}_\mathrm{2}$: seismic methods, gravimetry, and electrical/EM methods~\cite{Geophysical-2011-Fabriol, Quantification-2011-Korre}. Among those methods, the seismic method is presently 
without any doubt the most powerful method in terms of plume mapping, quantification of the injected volume in the reservoir and early detection
of leakage~\cite{Geophysical-2011-Fabriol}. In this work, we employ seismic data to detect the leakage of the $\mathrm{CO}_\mathrm{2}$.

\begin{figure}
 \begin{center}
    \includegraphics[width=0.5\linewidth]{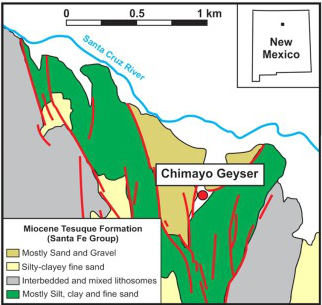}
  \end{center}
\caption{Map of Chimayó, New Mexico~\cite{Eruption-2014-Watson}. Faults are shown as red lines. The location of the geyser is shown as red dot.}
\label{fig:Chimayo_Geyser}
\end{figure}

The sedimentary basins of Chimayó, New Mexico have become prominent field laboratories for $\mathrm{CO}_\mathrm{2}$ sequestration 
analogue studies due to the naturally leaking $\mathrm{CO}_\mathrm{2}$ through faults, springs, and wellbores~\cite{Characteristics-2013-Han}.
The site of Chimayó Geyser is located in Chimayó, New Mexico within the Espanola Basin~(Fig.~\ref{fig:Chimayo_Geyser}). The bedrock consists predominately of sandstones cut by north–south trending faults. Chimayó geyser lies near the Roberts Fault and may cut directly through it. The source of $\mathrm{CO}_\mathrm{2}$ is unknown for the region. The regional aquifer supplying Chimayó geyser is semi-confined. The well was originally drilled in 1972 for residential water use but ended up tapping into a $\mathrm{CO}_\mathrm{2}$-rich water source and has geysered ever since. It has a diameter of 0.10~m, depth of 85~m and is cased with PVC for the entire depth. 

\begin{figure}
 \begin{center}
    \includegraphics[width=0.7\linewidth]{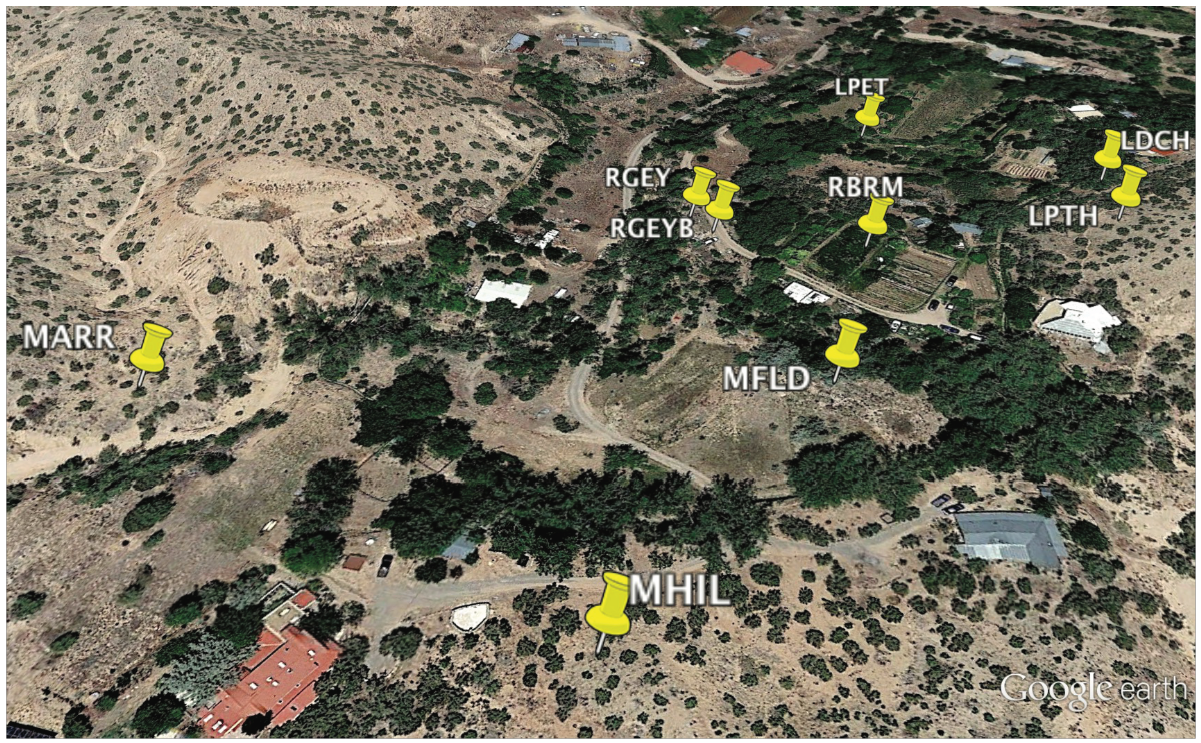}
  \end{center}
\caption{The distribution of seismic stations. The location of each seismic station is shown in yellow pin. }
\label{fig:station_distribution}
\end{figure}

To acquire seismic data, multiple stations are deployed at several points of interest, continuously recording time series signals as a time-amplitude representation. The locations of the seismic stations are shown in Fig.~\ref{fig:station_distribution}. The time series data has three components, representing amplitudes of three perpendicular directions. In this work, we only use one component of the signals from one station. 

\begin{figure}[t]
\centering
\subfloat[]{\includegraphics[width=0.49\linewidth]{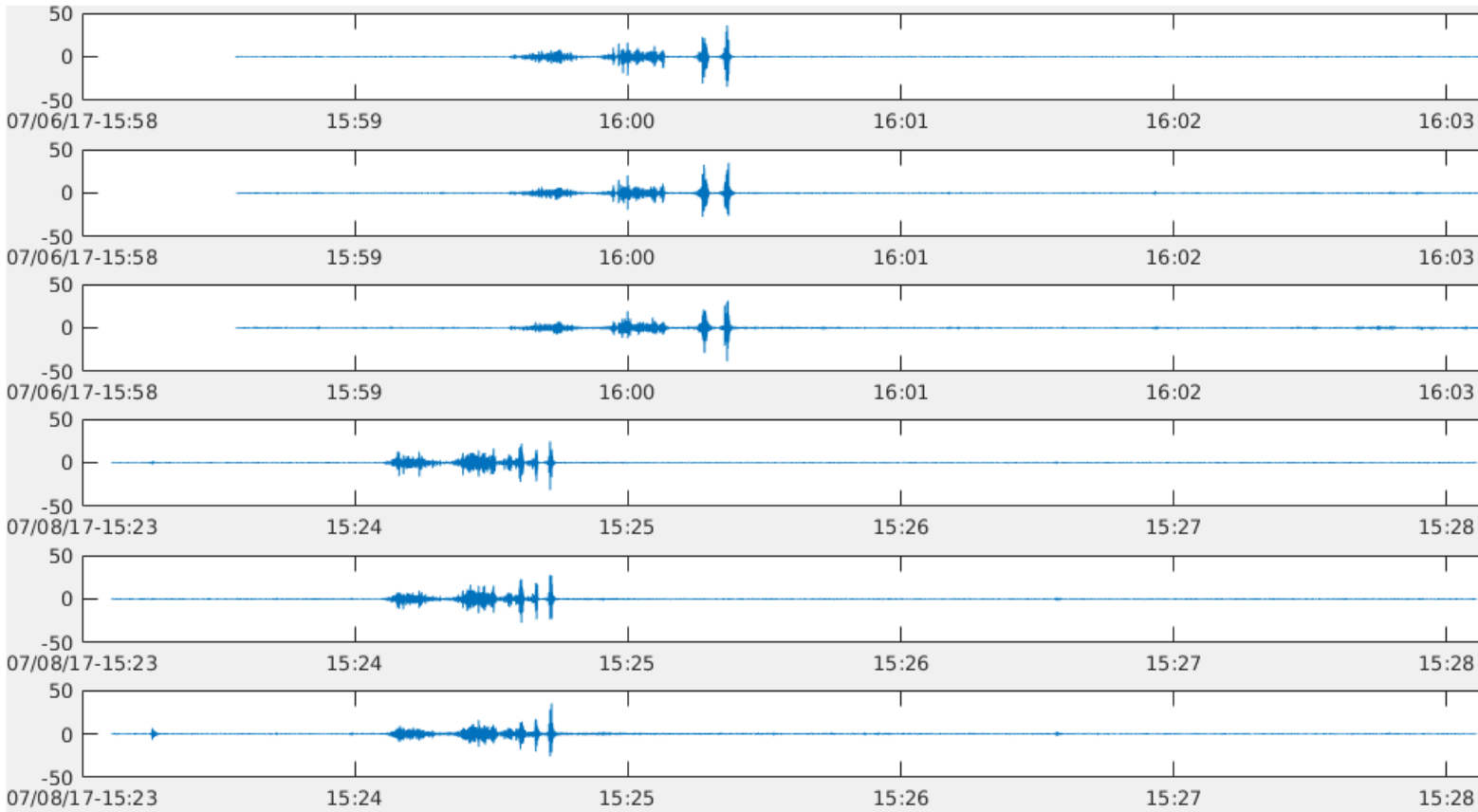}%
\label{fig:geyser_intro}}
\hfil
\subfloat[]{\includegraphics[width=0.41\linewidth]{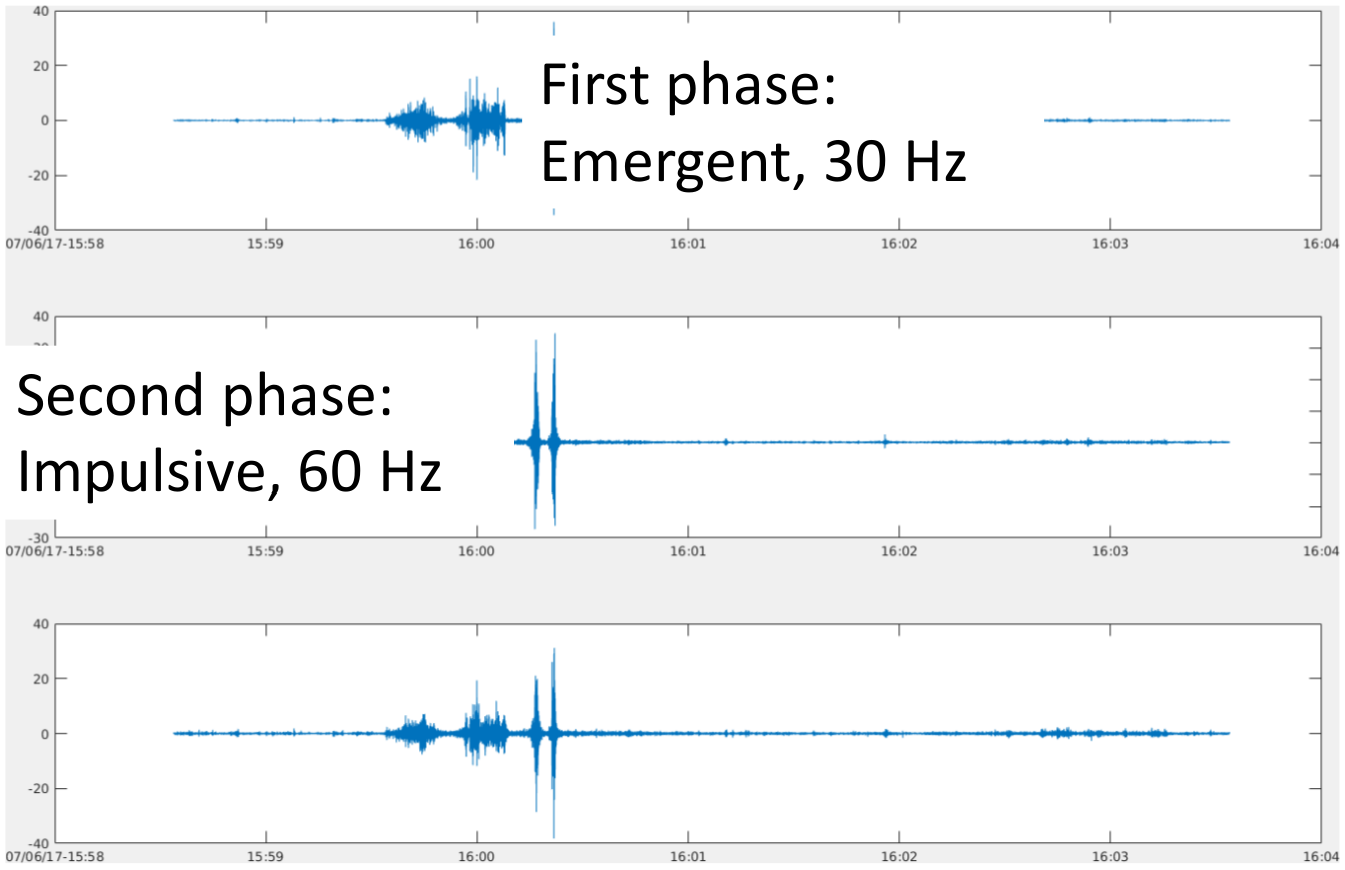}%
\label{fig:geyser_two_phase}}
\caption{(a): Time series signals of two geyser events, each with three perpendicular channels. 
(b): An illustration of the two phases of a geyser event. The first phase is called \textit{emergence} with 30 Hz of frequency. The second phase is called \textit{impulsive} with 60 Hz of frequency.}
\label{fig:geyser}
\end{figure}

Figure~\ref{fig:geyser_intro} illustrates the signals of two geyser events. Each event can be separated into two phases--\textit{emergent} and \textit{Impulsive}, which are shown in Fig.~\ref{fig:geyser_two_phase}. It is worthwhile to mention that the number of amplitude peaks in the second phase may be arbitrary. 

Geyser event detection from seismic data can be formulated as event detection problem with time series signals. Our main idea is to slide a fixed-length window through the time series data and use a binary classifier to detect whether there is a geyser event within. Tradition event detection algorithms that are widely used in the geophysics community are mainly similarity-based~\cite{TemplateMatching,autocorrelation-2008-Brown,Earthquake-2015-Yoon}. These algorithms are extremely inefficient (it takes weeks to iterate over the whole time series data) and not very accurate. Thus, we aim at developing machine learning models to significantly accelerate the inference procedure and boost the accuracy. 

Convolutional neural networks (CNN) have demonstrated the great potential to process image data~\cite{AlexNet,VGG}. It is natural to translate similar methodology into our 1D time series scenarios. CNN is particularly powerful to capture target patterns. For time series data, local dynamics are captured by shallow layers of a CNN due to the local connectivity. As the network becomes deeper, longer-range dynamics can be captured with the increasing size of the receptive field, that is, the pattern of the target waveform can be captured with a deep CNN. 

The state-of-the-art CNN architectures have been revolutionized since~\citet{ResNet,ResNet(b)}, known as ResNet where skip connections are built to make the network substantially deep. The merit of skip connections can be understood from two perspectives. Firstly, they enable the gradient from the output layer backpropagate directly to shallow layers, alleviating the notorious gradient vanishing problems~\cite{LeCun89}. 

Secondly, the whole architecture can be viewed as a huge ensemble since there are numerous paths from the input layer to the output layer, which significantly increases the robustness of the model and reduces overfitting. In this work, we built our model upon the densely connected block~\cite{DenseNet} (DenseNet), which is an improved version of ResNet. In a densely connected block, the output of a convolution layer is \textit{densely connected} with all previous outputs. The ``connection'' is implemented by concatenation operation. Thus, it fully exploits the advantage of the skip connection, while keeping a reasonable number of parameters through the reuse of features.

In this work, we explore the potential of the densely connected network in time series classification tasks. We use geyser data collected by several stations to evaluate our model. The experimental results show that our model can accurately detect geyser events without bells and whistles, which suggests the potential of the densely connected network for addressing challenging time series tasks. 

\section{Proposed Model: Seismic-Net}

In this section, we describe the structure of our Seismic-Net. 

\subsection{Densely Connected Block}
A densely connected block is formulated as
\begin{align}
x_{l + 1} & = \mathcal{H}([x_{0}, x_{1}, ..., x_{l}]) \\
\mathcal{H}(x) & = W*(\sigma(B(x))),
\label{eq:denseblock}
\end{align}
where $W$ is the weight matrix, the operator of ``*'' denotes convolution, $B$ denotes batch normalization (BN)~\citep{BN}, $\sigma(x) = \text{max}(0, x)$ ~\citep{ReLU} and $[x_{0}, x_{1}, ..., x_{l}]$ denotes the concatenation of all outputs of previous layers. 

The feature dimension $d_{l}$ of $x_{l}$ is calculated as
\begin{equation}
d_{l} = d_{0} + k \cdot l,
\end{equation}
where k, the \textit{growth rate}, is the number of filters used for each convolution layer.

\begin {table}
\centering
\begin{tabular}{ |c|c|c| }
\hline
Stage & Layers & Dim.\\
\hline
Input & - & 18000 $\times$ 1 \\
\hline
Convolution & conv7, 24, /2 & 9000 $\times$ 24 \\
\hline
Pool & avg-pool2, /2 & 4500 $\times$ 24 \\
\hline
$D_1$ & [conv3, 12] $\times$ 6 & 4500 $\times$ 96 \\
\hline
Pool & avg-pool2, /2 & 2250 $\times$ 96 \\
\hline
$D_2$ & [conv3, 12] $\times$ 6 & 2250 $\times$ 168 \\
\hline
Pool & avg-pool2, /2 & 1125 $\times$ 168 \\
\hline
$D_3$ & [conv3, 12] $\times$ 6 & 1125 $\times$ 240 \\
\hline
Pool & avg-pool2, /2 & 563 $\times$ 240 \\
\hline
$D_4$ & [conv3, 12] $\times$ 6 & 563 $\times$ 312 \\
\hline
Pool & avg-pool2, /2 & 282 $\times$ 312 \\
\hline
$D_5$ & [conv3, 12] $\times$ 6 & 282 $\times$ 384 \\
\hline
Pool & avg-pool2, /2 & 141 $\times$ 384 \\
\hline
$D_6$ & [conv3, 12] $\times$ 6 & 141 $\times$ 456 \\
\hline
Pool & avg-pool2, /2 & 72 $\times$ 456 \\
\hline
$D_7$ & [conv3, 12] $\times$ 6 & 72 $\times$ 528 \\
\hline
Pool & avg-pool2, /2 & 36 $\times$ 528 \\
\hline
$D_8$ & [conv3, 12] $\times$ 6 & 36 $\times$ 600 \\
\hline
Pool & avg-pool2, /2 & 18 $\times$ 600 \\
\hline
$D_9$ & [conv3, 12] $\times$ 6 & 18 $\times$ 672 \\
\hline
Pool & avg-pool2, /2 & 9 $\times$ 672 \\
\hline
$D_{10}$ & [conv3, 12] $\times$ 6 & 9 $\times$ 744 \\
\hline
Pool & avg-pool9, 9 & 1 $\times$ 744 \\
\hline
\multicolumn{3}{|c|}{1-d fully connected, logistic loss} \\
\hline
\end{tabular}
\caption{Network architecture of our Seismic-Net. This model is designed for inputs with 18,000 timestamps, which is ideal to effectively capture all geyser events.}
\label{table:densenet_architect}
\end {table}

\subsection{Network Architecture}

Table~\ref{table:densenet_architect} illustrates the overall architecture of our network. All convolution kernels in our network are 1 dimensional because of the input of 1D time series data. Conv7, 64, /2 denotes using 64 $1 \times 7$ convolution kernels with stride 2. The same routine applies to pooling layers. L denotes the length of input waveform. The brackets denote densely connected blocks, formulated in Eq.~\eqref{eq:denseblock}. We set the growth rate $k=12$ for all densely connected blocks, which results in $744$ extracted features for one time series segment. All densely connected blocks are followed by an average pooling layer, which downsamples the signals by 2. The last average pooling layer averages features over all timestamps to capture global dynamics. The global average pooling layer is followed by a fully connected layer, which is implemented as 
\begin{equation}
	x^{(l+1)} = W\cdot x^{(l)} + b,
\end{equation}
where $W$ is the weight matrix and $b$ is the bias. We use logistic loss as the loss function, which has the form 
\begin{equation}
	\text{LL}(y, z) = \text{log}(1 + e^{-yz}),
\end{equation}
where $y\in \{-1, 1\}$ is the ground-truth, $z$ is the predicted score. In the inference stage, the predicted label is based on the sign of the score.  

This architecture is designed for inputs that have 18,000 timestamps ($L=18,000$). This size of input signal is carefully chosen to effectively cover all geyser events. 

\section{Implementation Details}
In the training stage, we take 18,000-timestamp segments with geyser event included as positive samples. Negative samples includes manually picked 83 segments and 330 randomly picked segments from non-event signals. Since we only have the geyser event annotated, we found that manually picking most representative negative events necessary to achieve a good performance. We pre-process the time series data by substracting the mean, then dividing the standard deviation. We use Adam optimizer~\cite{Adam}, a variant of stochastic gradient descent algorithm, to minimize the loss function. We set the batch size to 50. The total number of parameters of our model is approximately 800K. It takes roughly 50 epochs to converge. Our model is implemented in TensorFlow~\cite{tensorflow}. We trained our model on a single NVIDIA GTX 1080 GPU. 

To test our model, we feed 18,000-timestamp windows into the model with 6,000-timestamp offset. We store the score of each positive detections. For multi-detections of the same event, we only keep the one with the highest score and discard the rest.  

\section{Experimental Results}
\subsection{Seismic Data}

We deploy multiple stations to continuously recording time series signals near the point of interests. The distribution of the stations are shown in Fig.~\ref{fig:station_distribution}. In this work, we only use signals from RGEYB to train and test our model. We have labeled signals in 55 days. Signals from 23 days, including 33 events, are used for training. Signals from 32 days, including 26 events, are used for testing. A day-long data has 17,280,001 timestamps. Geyser eruption happens at most twice a day. The longest geyser event in our dataset spans 12,000 timetamps. 

\begin{figure}
\centering
\includegraphics[width=0.9\linewidth, height=5cm]{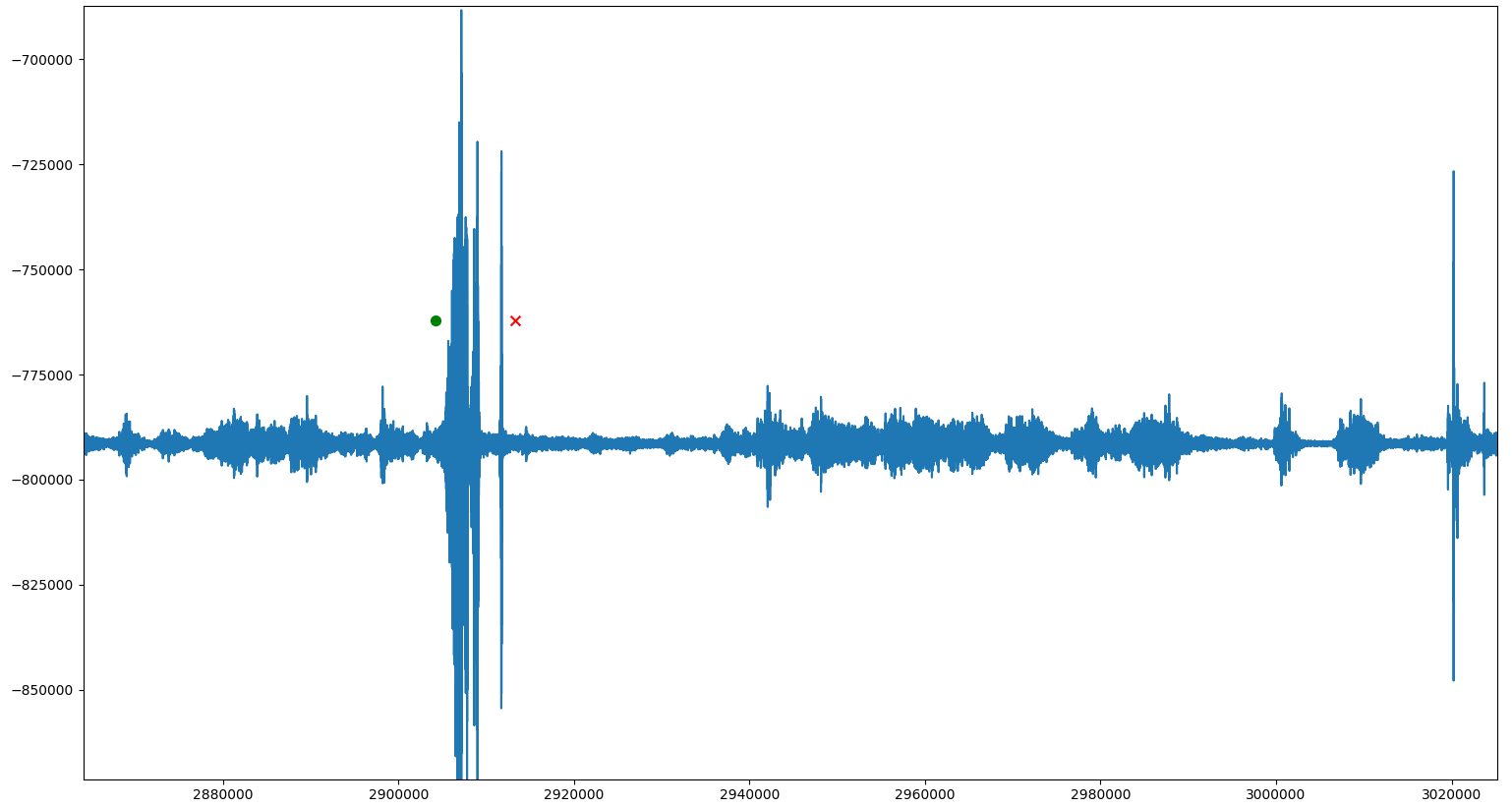}
\caption{A segment of our geyser data. A geyser event is in the window indicated by a green dot and red cross. }
\label{fig:geyser_illu}
\end{figure}

Figure~\ref{fig:geyser_illu} gives an illustration of a segment of our data with a geyser event included. The event happens within the window bounded by a green dot and a red cross. All other signals are considered noise. As previously mentioned, some negative samples are hand-picked due to the unknown negative sample space. Some negative samples include eruptions caused by passing trains and human activities, which may fool the classifier if not included in the training set. 

\subsection{Results}

\begin{table}[ht]
\centering
\begin{tabular}{ c|c|c|c }
& Precision & Recall & \# Parameters \\ 
\hline
Kernel SVM & fail & fail & 18K \\
\hline
VGG-based Net & 0.609 & 0.538 & 37M \\
\hline
ResNet-based Net & 0.581 & 0.960 & 15M \\
\hline
Seismic-Net & \textbf{0.889} & \textbf{0.923} & 800K \\
\end{tabular}
\caption{The detection results given by kernel SVM, CNN inspired by VGG net, ResNet-based CNN and densely connected CNN. ``fail'' indicates extremely low accuracy. These results indicates that the shallow model~(SVM) is greatly outperformed by deep models. The densely connected network stands out among other CNN-based models.}
\label{table:results}
\end{table}

The detection results w.r.t. several different methods are provided in Table~\ref{table:results}. We compare the performances of four models: 1) support vector machine with radial basis function kernel; 2) VGG-based~\cite{VGG} CNN; 3) ResNet-based CNN and 4) DenseNet-based CNN. We use the same training routine as the proposed CNN to train the kernel SVM. We found that the kernel SVM has extremely low accuracy so we put ``fail'' in the table. It suggests that advanced feature extraction techniques are required in advance to apply SVM, logistic regression or other shallow models. 

The VGG-like network has no skip connections, which is the major difference with ResNet-based or DenseNet-based network. Since the input size is large (18,000 timestamps), more convolution layers are required to capture the global dynamics, which makes the number of parameters in VGG-based network substantially. The optimization for VGG-based network is rather difficult without the help of skip connections. Thus, it has the lowest accuracy among the three CNN-based models. The ResNet-based model, with skip connections, has the second best performance. The DenseNet-based model, with densely connected convolution layers, has the best performance. It has 24 correct detections (TP) out of 27 (TP + FP) total detections. The number of events in test set is 26 (TP + FN). We use precision and recall as the evaluation metrics. They are calculated as
\begin{align}
\text{Precision} = \frac{\text{TP}}{\text{TP} + \text{FP}}, \\
\text{Recall} = \frac{\text{TP}}{\text{TP} + \text{FN}},
\end{align}
where TP stands for ``true positive'', FP stands for ``false positive'' and FN stands for ``false negative''.

\section{Example Detections}
\begin{figure*}[ht]
\centering
\subfloat[]{\includegraphics[width=0.4\linewidth]{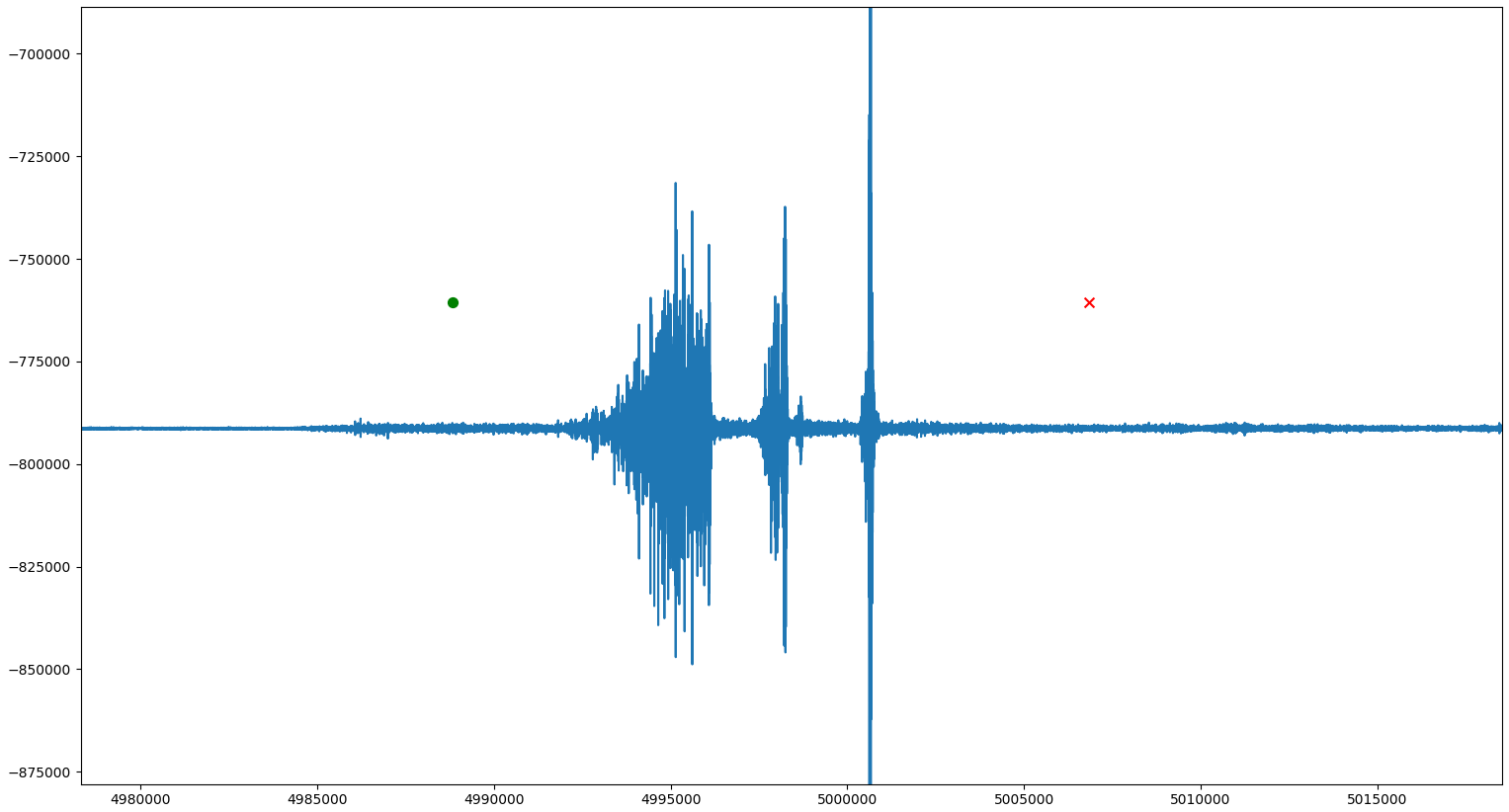}}
\hfil
\subfloat[]{\includegraphics[width=0.4\linewidth]{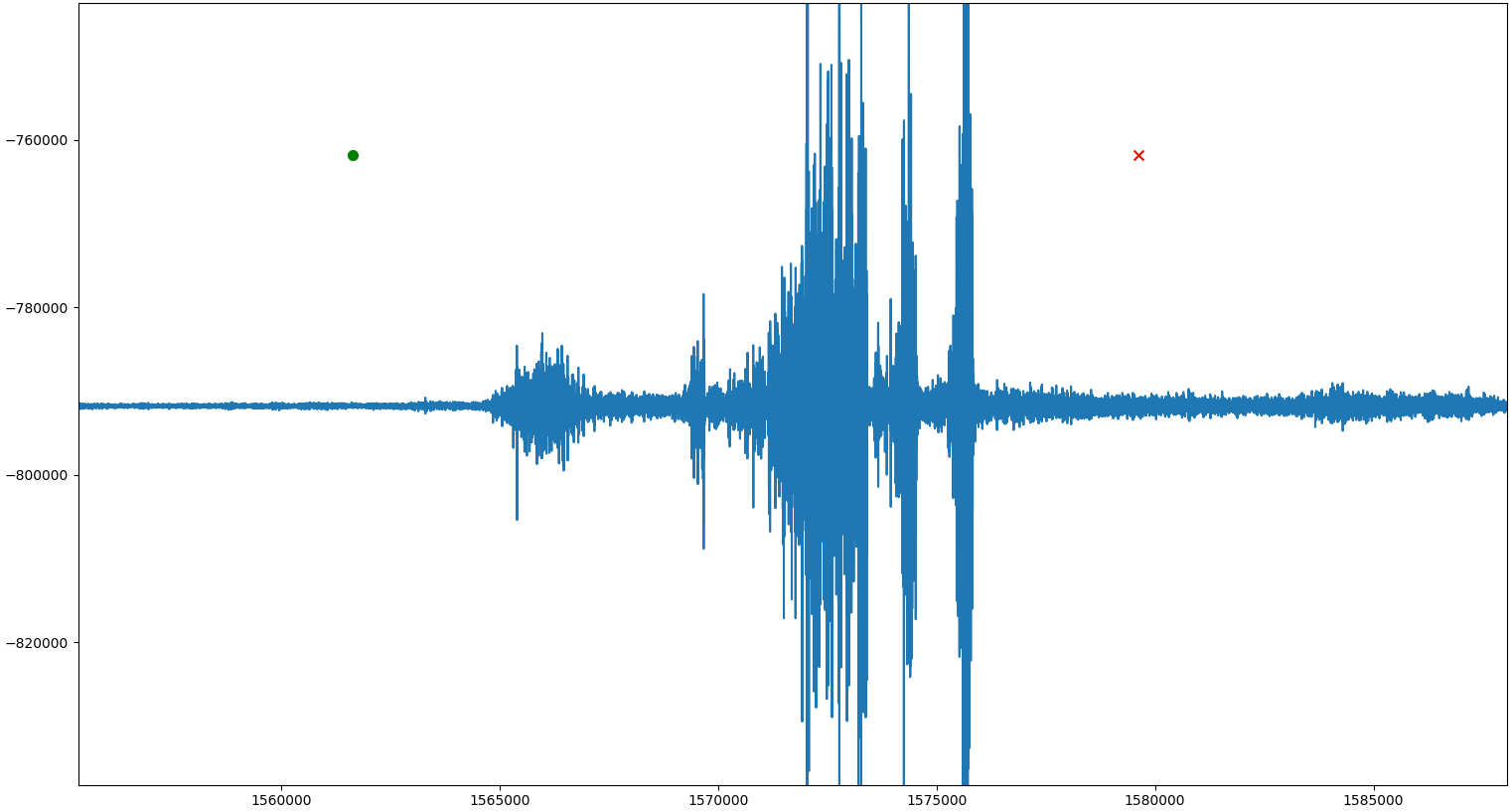}}
\hfil
\subfloat[]{\includegraphics[width=0.4\linewidth]{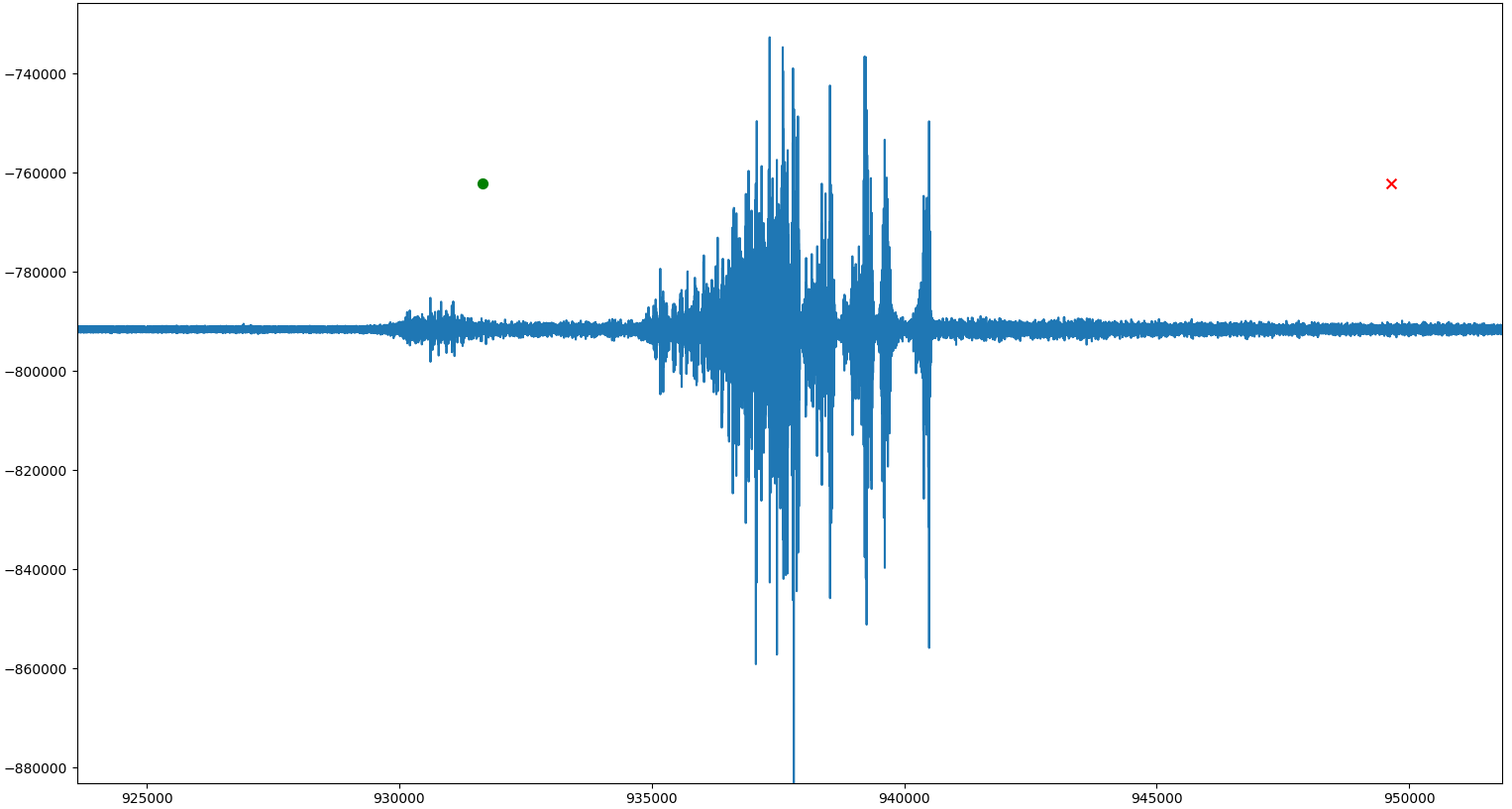}}
\hfil
\subfloat[]{\includegraphics[width=0.4\linewidth]{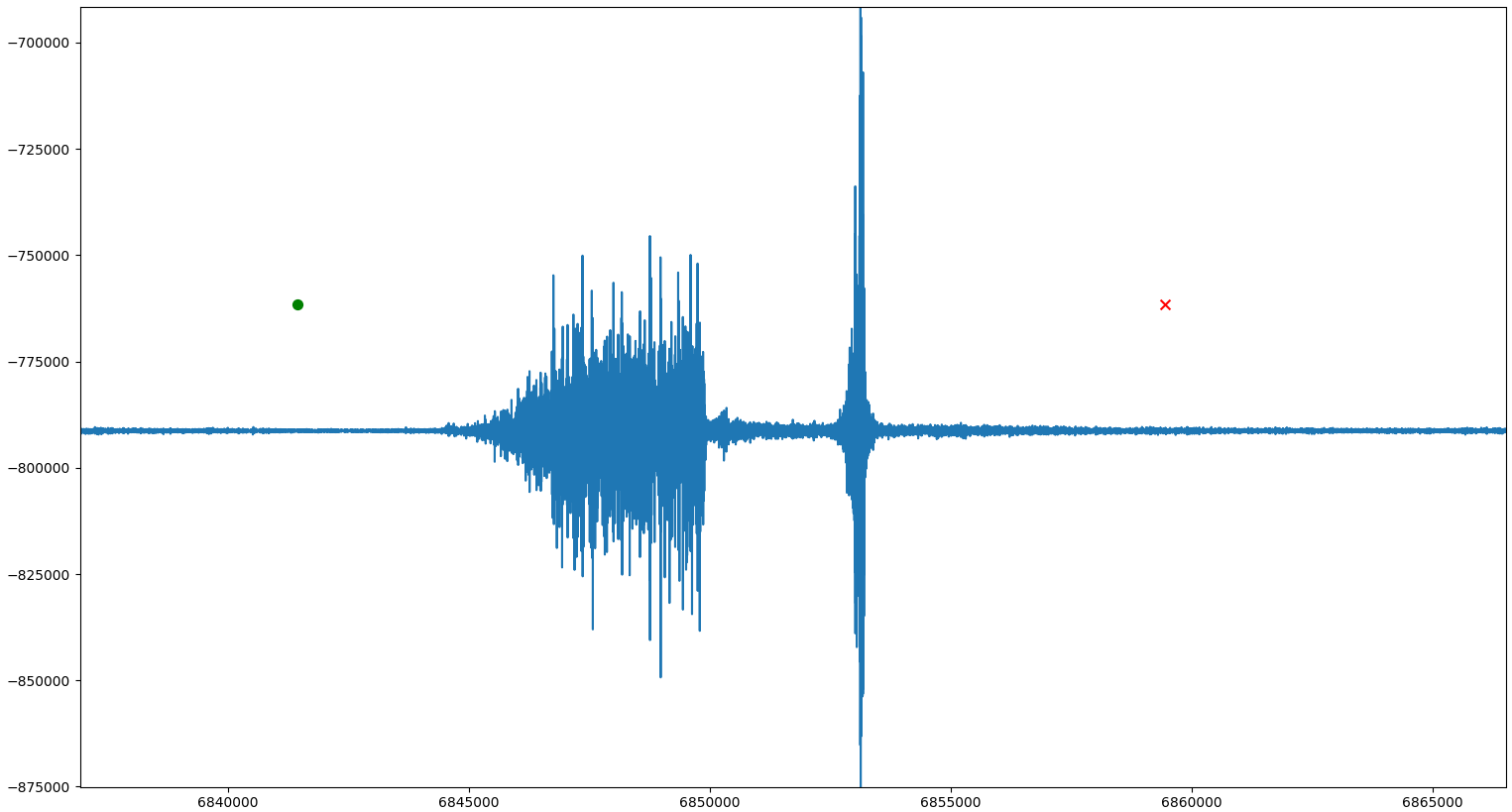}}
\caption{Some example detections are shown here. These four events demonstrate that our model is able to accurately detect geyser events even though the patterns are not always similar since the number of eruptions in the two phases is arbitrary. The green dot and the red cross indicate the location of the sliding window where our model gives positive predictions. }
\label{fig:example_detection}
\end{figure*}

We present four detections given by our model in Fig~\ref{fig:example_detection}. These results indicate the potential of our model to accurately capture specific patterns in time series signals. Although the patterns of geyser events vary, which is caused by the arbitrary number of eruptions in the second phase, those events are still correctly detected by our DenseNet-based CNN. The waveforms of the second phase are those ``thinner'' eruptions. Fig.~\ref{fig:example_detection} (a),
Fig.~\ref{fig:example_detection} (b),
Fig.~\ref{fig:example_detection} (c),
 and Fig.~\ref{fig:example_detection} (d) have 2, 2, 4 and 1 amplitude peaks in the second eruption, respectively. The locations of the triggered sliding window is indicated by a green dot and a red cross. 
 
\section{Conclusion}

In this work, we developed a seismic event detection package, entitled ``Seismic-Net''. Our Seismic-Net is based on convolutional
neural network and is  capable of detecting seismic events accurately and efficiently. Our network is substantially deep in order to capture global dynamics. Densely connected blocks are inserted to reduce the number of parameters and simplify the optimization. We employ our Seismic-Net 
to a natural analog site at Chimayó, New Mexico. The experiment results demonstrate that our model achieves high accuracy without bells and whistles. By comparing with shallow models and other convolutional neural network  variants, we justify that the proposed DenseNet-based architecture is the most accurate model. It is also the most efficient among convolutional neural network-based models with only 800K parameters. Thus, we conclude the proposed model has a great potential capture complicated patterns in time series data, therefore can be used for an early detection of $\mathrm{CO}_\mathrm{2}$ leakage.

\section*{Acknowledgment}
This work was co-funded by the Center for Space and Earth Science at Los Alamos National Laboratory, and the U.S. DOE Office of Fossil Energy through its Carbon Storage Program.

\bibliographystyle{IEEEtranN}
\bibliography{bibliography}

\end{document}